\documentclass[conference]{IEEEtran}
\IEEEoverridecommandlockouts
\usepackage{cite}
\usepackage{amsmath,amssymb,amsfonts}
\usepackage{algorithm}
\usepackage{array}
\usepackage{algpseudocode}
\usepackage{graphicx}
\usepackage{hyperref}
\usepackage{url}
\usepackage{textcomp}
\usepackage{xcolor}
\def\BibTeX{{\rm B\kern-.05em{\sc i\kern-.025em b}\kern-.08em
    hT\kern-.1667em\lower.7ex\hbox{E}\kern-.125emX}}
\begin{document}

\title{Towards Self-Adaptive Machine Learning-Enabled Systems Through QoS-Aware Model Switching}

\author{\IEEEauthorblockN{Shubham Kulkarni, Arya Marda, Karthik Vaidhyanathan}
\IEEEauthorblockA{\textit{Software Engineering Research Center},  \textit{IIIT Hyderabad}, India\\
shubham.kulkarni@research.iiit.ac.in, arya.marda@students.iiit.ac.in, karthik.vaidhyanathan@iiit.ac.in
}}

\maketitle
\begin{abstract}
Machine Learning (ML), particularly deep learning, has seen vast advancements, leading to the rise of Machine Learning-Enabled Systems (MLS). However, numerous software engineering challenges persist in propelling these MLS into production, largely due to various run-time uncertainties that impact the overall Quality of Service (QoS). These uncertainties emanate from ML models, software components, and environmental factors. Self-adaptation techniques present potential in managing run-time uncertainties, but their application in MLS remains largely unexplored. As a solution, we propose the concept of a Machine Learning Model Balancer, focusing on managing uncertainties related to ML models by using multiple models. Subsequently, we introduce AdaMLS, a novel self-adaptation approach that leverages this concept and extends the traditional MAPE-K loop for continuous MLS adaptation. AdaMLS employs lightweight unsupervised learning for dynamic model switching, thereby ensuring consistent QoS. Through a self-adaptive object detection system prototype, we demonstrate AdaMLS's effectiveness in balancing system and model performance. Preliminary results suggest AdaMLS surpasses naive and single state-of-the-art models in QoS guarantees, heralding the advancement towards self-adaptive MLS with optimal QoS in dynamic environments.
\end{abstract}


\section{Introduction}
Recent advancements in machine learning, especially deep learning, have spurred the growth of Machine Learning-Enabled Systems (MLS) like ChatGPT~\cite{chatgpt}, Google Bard~\cite{bard}, and DALLE-2~\cite{dalle2}. However, engineering MLS presents multifaceted software engineering challenges, from the development and integration of ML components to model versioning and data quality~\cite{graceml, henrykarthikml, microsoft}. Gartner's report notes that almost half of MLS projects don't make it to production, mainly because of unpredictable run-time challenges like varying model performance and unstable software components~\cite{Gartner}. Additionally, environmental factors like infrastructure (cost, energy) and system usage (arrival rate etc.) significantly influence the system QoS. Over the years, self-adaptation techniques have emerged as promising solutions for managing run-time uncertainties ~\cite{weyns, selfchapter}. They enable systems to continuously adapt their structure and/or behaviour to satisfy different goals (in terms of QoS, functionalities, etc.). While effective in domains like CPS, IoT, and service-oriented systems~\cite{bradley, qvml, tele}, their application in MLS is largely unexplored~\cite{garlan}. 
In MLS, ML model performance can vary significantly due to factors like model architecture—layer count and algorithm type. Given identical input-output specifications, developers can devise a spectrum of models, each with its speed and accuracy trade-offs. Recognizing this variability, we introduce the concept of an ML Model Balancer. This notion encapsulates the idea of dynamically evaluating and switching between models to optimize QoS. For instance, high-traffic situations might favor a faster model, while quieter periods prioritize accuracy. 
AdaMLS, our novel self-adaptive approach, operationalizes this concept of the ML Model Balancer. Nevertheless, AdaMLS consistently excels in navigating the intricacies of online ML deployments, ensuring superior QoS. This includes: i) monitoring model and system parameters; ii) analyzing model and system quality for QoS violations; iii) using knowledge from lightweight unsupervised learning to dynamically switch models, ensuring QoS; and iv) executing system adaptation. Prioritizing ML model adaptability, AdaMLS shifts from conventional load balancing to QoS-aware dynamic ML model switching. By continuously tuning model selections in response to environmental cues and system demands, AdaMLS guarantees MLS QoS, promoting consistent MLS operation in live settings. This represents a stride towards future-ready self-adaptive MLS, designed to maintain an optimal performance equilibrium amidst changing data and user demands. 
We evaluate AdaMLS using an object detection use case through utility (refer section \ref{sec:Running_example}) showcasing a self-adaptive prototype. Our preliminary findings indicate that the runtime model switching, facilitated by lightweight unsupervised learning, effectively manages both system and model performance. This enables AdaMLS to surpass both naive strategies and individual models in terms of Quality of Service (QoS). Our work innovatively adapts the MAPE-K loop to address the uncertainties inherent in MLS, emphasizing dynamic model-switching approach. Through AdaMLS's real-world application, we highlight our move toward self-adaptive MLS that can deftly switch between models based on data shifts and user demands, always maintaining optimal QoS. The paper is structured as follows: Section 2 provides motivation with a running example. Section 3 introduces the AdaMLS approach. Results from its application are in Section 4. Related work is discussed in Section 5, and Section 6 concludes.
\begin{table*}[htbp]
\centering
\caption{Reducing Uncertainty in Self-adaptive Systems}
\begin{tabular}{|m{2.5cm}|m{4.0cm}|m{7.8cm}|m{2cm}|}
\hline
\textbf{Uncertainty Source} & \textbf{Uncertainty Explanation} & \textbf{Mitigation Strategy} & \textbf{Implemented In}\\
\hline
ML Models in MLS & Abstraction of ML model & Evaluate all models on test set; abstract performance with statistics & Learning Engine \\
\hline
Resources & Dynamicity in the system & Preload ML Models; reduce latency and adaptation action costs & MLS \\
\hline
Environment & Incoming requests Unpredictability & Monitor request rate and produced results continuously & Monitor \\
\hline
Adaptation Function & Impact of tactic & Utilize confidence intervals for adaptation rules, ensuring assurance & Learning Engine \\
\hline
Goal & Conflicting or dependent goals & Prioritize most accurate model among eligible models for requests & Planner \\
\hline
Data Drift & Untrained or rare data & Map performance for past 'n' requests to nearest cluster for insight & Analyzer \\
\hline
Model Drift & Model performance decay & Rank models per goal; auto-remove degraded models & Planner \\
\hline
\end{tabular}
\label{tab1}
\end{table*}

\section{Running Example}
\label{sec:Running_example}
Our AdaMLS\footnote{unless specified otherwise by model we imply ML model in this paper} approach is showcased via an object detection system, a culmination of ML advancements over decades~\cite{obj20}. The system consists a {\em {web service}} with a REST API, {\em model\_repo} as the repository, {\em message\_broker} for image streaming, and {\em obj\_model} using YOLO~\cite{yolo}. These components mirror services like Google Cloud Vision or Amazon Rekognition, emphasizing real world applicability.
In the example, we define a set $M$ of available models. Each model $m_j$, where $j$ ranges from 1 to $n$, is part of $M$. Here this set includes YOLOv5 models (YOLOv5n, YOLOv5s, YOLOv5m, YOLOv5l, YOLOv5x) provided by Ultralytics~\cite{yolov5}, pretrained on the COCO 2017 training dataset~\cite{coco}. Models in $M$ are quantified as mAP - model's effectiveness in detecting objects, symbolized by $c$, and performance. Performance is assessed using $\tau'$, $\tau$, and $r$. Here, $\tau'$ denotes the processing time per image by system (i.e., individual processing without network or queuing delay), $\tau$ is the model's processing time, and $r$ is the system response time in real-world operations, encompassing network, queuing, and processing delays. For instance, YOLOv5n has 1.9M parameters, an mAP of 28, and a 45-ms $\tau$, while YOLOv5x has 86M parameters, mAP 50.7, and 766 ms $\tau$~\cite{yolov5}.
Different models vary in response time and confidence scores, with none achieving an optimal balance between both.
Given this context, and given a set of thresholds including $C_{\max}$ and $C_{\min}$, denoting the maximum \& minimum confidence score; $R_{\max}$ and $R_{\min}$, the maximum \& minimum allowed response time; the goal is to maximize the utility function $U$. This function evaluates the confidence score $c_i$ and response time $r_i$ for each image $i$. Herein, $p_{ev}$ and $p_{dv}$ represent the penalties for violations relative to these thresholds. The total utility $U$ of the system for all $k$ unique image ID processed, is given by $U = \sum_{i=1}^{k} U_i$. For the $i^{th}$ image the utility $U_i$ is defined as $U_i = w_e E_{\tau_i} + w_d T_{\tau_i}$, where $w_e$ and $w_d$ are weights, $E_{\tau_i}$ and $T_{\tau_i}$ are piece-wise functions that represent the $c$ and $r$ respectively, defined as:

\begin{scriptsize}
\[
  E_{\tau_i} =
  \begin{cases}
        c_i & \text{if } C_{\min} \leq c_i \leq C_{\max} \\
        (c_i - C_{\max})\cdot p_{ev} & \text{if } c_i > C_{\max} \\
        (C_{\min} - c_i)\cdot p_{ev} & \text{if } c_i < C_{\min}
  \end{cases}
\]

\[
  T_{\tau_i} =
  \begin{cases}
        r_i & \text{if } R_{\min} \leq r_i \leq R_{\max} \\
        (R_{\max} - r_i)\cdot p_{dv} & \text{if } r_i > R_{\max} \\
        (r_i - R_{\min})\cdot p_{dv} & \text{if } r_i < R_{\min}
  \end{cases}
\]
\end{scriptsize}

Given the thresholds and constraints, our approach aims to maximize utility function thereby improving the overall QoS.

\section{AdaMLS Approach }
AdaMLS provides a robust solution to two pivotal learning problems\cite{danny}: adaptation policy development and resource usage analysis.Leveraging unsupervised learning to identify unknown patterns in MLS's runtime performance, the Learning Engine (LE) provides adaptation rules. As outlined in Table~\ref{tab1} and defined in~\cite{uncertanity} and ~\cite{garlan}, these rules, when executed by MAPE-K, effectively mitigate uncertainties.
\begin{figure}[htbp]
\centerline{\includegraphics[width=\columnwidth]{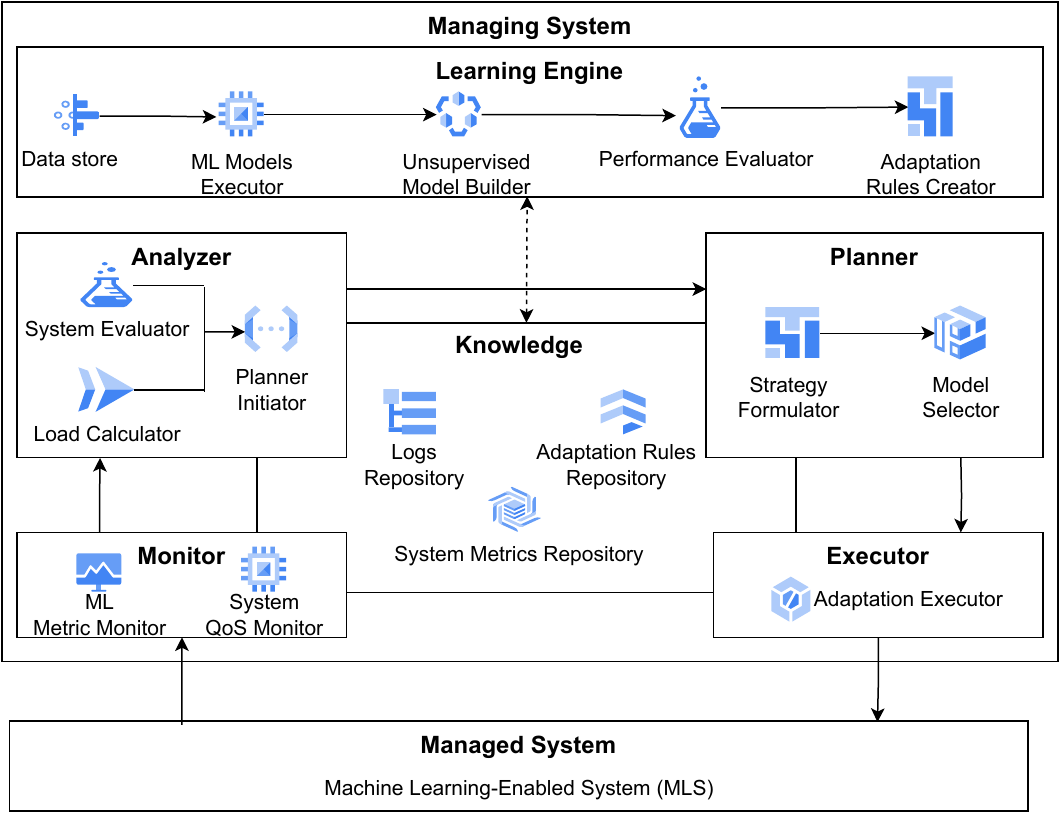}}
\caption{AdaMLS Approach}
\label{fig}
\end{figure}

\subsection{Learning Engine}
The Learning Engine (LE) initializes with the {\em ML Models Executor}, operating all models $m_j \in M$ on an evaluation dataset (e.g., COCO Test 2017) from the {\em Data Store}. It saves detection outcomes in dataset $d_{j}$ for each model. Outputs include KPIs like $c$, $\tau$, $r$, and $s$; where $s$ signifies CPU Consumption(\%), aiding in uncertainty mitigation as said in Table \ref{tab1}.The {\em Unsupervised Model builder} uses K-Means clustering on each $d_{j}$ based on $\tau'$, grouping models by performance attributes, thereby hastening model selection during runtime adjustments. Clustering with $\tau'$ is strategic, given our hierarchical approach where response time is paramount. $\tau'$ isn't arbitrary but marks our primary metric. This ensures models first align with the vital response-time criteria before secondary metrics such as accuracy. This clustering navigates model spectrum effectively, grouping models with akin performance, facilitating dynamic switches to select from relevant QoS subsets. Each image $i$ of model $m_j$ gets a cluster label $l$.

The {\em Performance Evaluator} constructs a performance matrix for each model $m_j \in M$, by collating KPIs across models ($m_q \in M$ where $q \neq j$) for the same request. Let's consider model $m_j$, known as 'nano' (Yolov5n). For each image $i$, KPIs are collated across all models using the Image ID, resulting in a comprehensive matrix for 'nano'. Each row in this matrix represents an image, while columns correspond to KPIs from all models and the assigned cluster number from 'nano'.

The {\em Adaptation Rule Creator} calculates the 90\% confidence intervals (CI) for each cluster $l$ of every model $m_j$. These upper and lower limit intervals provide a statistically likely range for KPIs, thereby reducing uncertainty in system adaptations as per Table \ref{tab1}. To illustrate, consider model 'nano' as $m_j$. For each cluster $l$ in 'nano', CIs are calculated for all data points, using only those images within the same cluster. Simultaneously, the CIs for the same images are calculated from the performance metrics of all other models $m_k$. This process results in a CI matrix for 'nano', encapsulating potential performance variations across models. Repeating this for all models, LE produces a set of CI matrices. Each matrix maps out performance variations within each cluster of the respective model $m_j$. Through LE, executed periodically in batches, AdaMLS develops a holistic understanding of potential model performance shifts, enabling statistical predictions of system KPIs impacts due to model switching.

\subsection{MAPE-K Loop}
\subsubsection{Knowledge} As per the structure depicted in Figure \ref{fig}, the Knowledge (K) base in our system is primarily a repository divided into three sections: the {\em Log Repository}, the {\em Adaptation Rule Repository}, and the {\em System Metrics Repository}. The {\em Log Repository} stores system logs, including vital KPIs. For instance, in our running example, this would mean processing time per image by the system $\tau$, the system response time $r$, model confidence $c$, and CPU consumption $s$ for all processed requests $k$, as recorded by the MLS. The {\em Adaptation Rule Repository} houses the CI matrix generated by the LE, acting as a set of adaptation rules for the Planning phase. Lastly, the {\em System Metrics Repository} keeps track of various system metrics such as real-time incoming request rate per second denoted as $v$ and system logs if any.
\subsubsection{Monitor} The {\em ML Metric Monitor} component continuously tracks system QoS and KPIs. In our use case, this includes the avg. number of detection boxes in processed image $b$,$\tau$,$r$,$c$ and $s$ from the last $i$ processed requests denoted as $k'$, and for {\em System QoS Monitor} component, it is the current model in use denoted as $m'$ and $v$ to mitigate environmental uncertainty as per Table \ref{tab1}. The system also tracks the number of pending requests $i_{w}$, which provides insight into the workload. The monitored data is sent to the Analyze function for potential adaptations maintaining system self-awareness also logged in Knowledge appropriately.
\subsubsection{Analyzer} In Analyzer phase, the {\em Planner Initiator} utilizes the {\em System Evaluator} to analyze the data from the Monitor, to determine if a system adaptation is necessary. The {\em System Evaluator} identifies the closest cluster, $l$, from the CI matrix for $m'$, to the current system state. This state is defined as the mean of the most recent $k'$ (e.g. 50) results for $m'$. This identification process is grounded in the two KPIs exhibiting the highest variance. Process mitigates data drift uncertainty as outlined in Table \ref{tab1}. Upon identifying cluster $l$, a feasible request rate range [$v_{min}, v_{max}$] is determined from CI matrix of $m'$ and is computed using the inverse of the upper and lower confidence interval bounds for $\tau$ for $m'$. Subsequently, the {\em Load Calculator} computes the adjusted request rate $v_{adj}$ by adding $i_w$ (those exceeding $v_{max}$) to $v$.

If $v_{adj}$ not in [$v_{min}$, $v_{max}$], the {\em Planner Initiator} waits for a brief period $t_{wait}$ (e.g. 0.25 sec) to avoid unnecessary system adaptations and then initiates Planner with $v_{adj}$, $m'$, and $l$.

\subsubsection{Planner} During the Planning phase, the {\em Strategy Formulator} uses the output from the Analyzer and Knowledge base to devise an adaptation strategy. The strategy identifies potential models from $M$ that can accommodate $v_{adj}$ and belong to cluster $l$. The compatibility of a model is determined by comparing $v_{adj}$ with the inverse of lower value of CI bound for $\tau$; for the current model $m'$, the most recent $n$ results are used, whereas for other models in $M$, the CI matrix of $m'$ is referenced. The {\em Model Selector} then picks $m_{best}$, the model with the highest lower confidence interval value for $c$, thereby mitigating goal and model drift uncertainties as per Table \ref{tab1}. If $m' = m_{best}$, the Planner refrains from taking further action. Otherwise, the Planner signals a model switch to the Execution phase. The system persists with $m'$ if no suitable model is identified.
\subsubsection{Executor} In the Execution phase, the {\em Adaptation Executor} carries out the plan. If the Planner signals a model switch, i.e., $m_{best} \neq m'$, the system transitions to $m_{best}$. If no switch is signaled, the system persists with the current model. Both situations guarantee system autonomy, adaptability to changing conditions, and enhanced learning through updates logged in the Knowledge base.
\section{Preliminary Results}
\label{sec:results}
We implement AdaMLS on an object detection system (refer section \ref{sec:Running_example}) using YOLOv5 variants alongside FastAPI. For testing, we emulate a FIFA98 situation~\cite{fifa}, with up to 28 parallel requests/sec and a total of 25,000 requests. We employ the COCO 2017 unlabelled dataset \cite{coco} as our testing dataset, with the COCO 2017 test dataset as evaluation dataset. Our data clustering is facilitated through Python and PySpark's MLlib \cite{pyspark}, with the optimal clusters being defined by the elbow method \cite{elbow}. The complete specifics of our implementation, parameters and the ensuing results are detailed in \cite{ourproject}. In our evaluation, AdaMLS is evaluated against both the naive approach and individual YOLOv5 models. The naive approach transitions between models based on preset $v$ thresholds. Referring to the data in Table \ref{tab:performance}, it is clear that AdaMLS effectively decreases the average response time $r$ while simultaneously reducing the occurrence of penalties associated with response time and confidence. Parameters used for utility are: $p_{ev}$ = 1,$p_{dv}$ = 1,$C_{\max}$ = 1,$C_{\min}$ = 0.5, $R_{\max}$ = 1s, $R_{\min}$ = 0.1s, $t_{wait}$ = 0.25s and $k'$ = 50. A key aspect of our results is the utility metric, detailed in Table \ref{tab:your_label}. While utility, a measure quantifying the effectiveness and efficiency of a model in various operational contexts (as detailed in Section \ref{sec:Running_example}), offers an effective measure, it's not the sole criterion for evaluating QoS in ML systems. Although AdaMLS does not consistently lead in all individual metrics, it demonstrates unparalleled efficacy in terms of utility. Specifically, AdaMLS achieves an overall increased utility, particularly when equal emphasis is placed on response time and confidence score, surpassing the Yolov5n model by as much as 39\%. This remarkable performance in utility, even when not consistently leading in every individual metric, is significant. It underscores our method's ability to integrate these metrics effectively for optimal outcomes. Therefore, utility acts as a measure of our system's proficiency in addressing challenges and maintaining high-quality performance. Moreover, our refined architecture reduces the time required for model transitions, ensuring it remains below the crucial threshold of 0.01 seconds, further solidifying AdaMLS's contribution to improving the QoS of ML-driven systems.

\begin{table}[htbp]
  \centering
  \caption{Performance Comparison}
    \begin{tabular}{|l|c|c|c|}
    \hline
    \textbf{Parameter} & \textbf{Yolov5n} & \textbf{Naive} & \textbf{AdaMLS} \\
    \hline
    Model Switch Instances & N/A & \textbf{49} & 308 \\
    Avg. Confidence Score (c) & 0.535 & \textbf{0.606} & 0.6 \\
    Avg. Response time (r) & \textbf{0.28} & 1.94 & 0.47 \\
    Avg. CPU Consumption & \textbf{42.94} & 66.42 & 65.27 \\
    Penalty Instances for r & \textbf{365} & 7731 & 927 \\
    Penalty Instances for c & 9770 & \textbf{5843} & 6137 \\
    \hline
    \end{tabular}%
  \label{tab:performance}%
\end{table}%

\begin{table}[h]
    \centering
    \caption{Utility Comparison}
    \scriptsize 
    \setlength{\tabcolsep}{0.2pt} 
        \begin{tabular}{|*{9}{c|}}
        \hline
        $w_e$ & $w_d$ & Yolov5n & Yolov5s & Yolov5m & Yolov5l & Yolov5x & Naive & AdaMLS \\ 
        \hline
        \textbf{0} & \textbf{1} & 5817 & $-7.9\times10^{5}$ & $-10\times10^6$ & $-39\times10^6$ & $-81\times10^6$ & -27313 & \textbf{9978} \\ 
        \hline
        \textbf{0.25} & \textbf{0.75} & 6487 & $-5.9\times10^{5}$ & $-7.8\times10^{6}$ & $-29\times10^{5}$ & $-61\times10^{6}$ & -17427 & \textbf{8992} \\ 
        \hline
        \textbf{0.5} & \textbf{0.5} & 7157 & $-3.9\times10^{5}$ & $-5.2\times10^{6}$ & $-19\times10^{6}$ & $-40\times10^{6}$ & -7541 & \textbf{9978} \\ 
        \hline
        \textbf{0.75} & \textbf{0.25} & 7827 & $-1.8\times10^{5}$ & $-2.5\times10^{6}$ & $-9.9\times10^{6}$ & $-20\times10^{6}$ & 2345 & \textbf{10964} \\ 
        \hline
        \textbf{1} & \textbf{0} & 8498 & 12566 & 14612 & 15270 & \textbf{16264} & 12231 & 11949 \\ 
        \hline
        \end{tabular}
        \label{tab:your_label}
\end{table}

\begin{figure}[htbp]
\centerline{\includegraphics[width=\columnwidth]{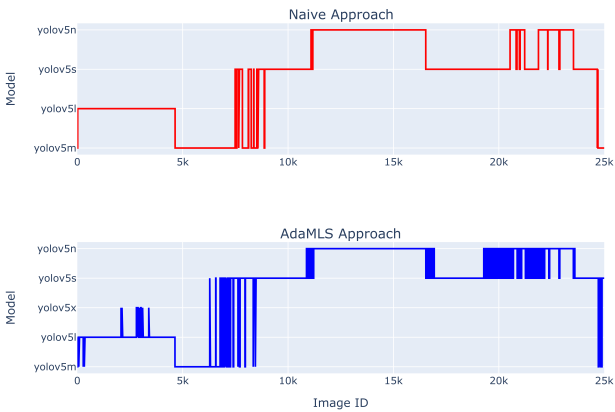}}
\caption{Model Switching: Naive Vs. AdaMLS}
\label{fig2}
\end{figure}

\begin{figure}[htbp]
\centerline{\includegraphics[width=\columnwidth]{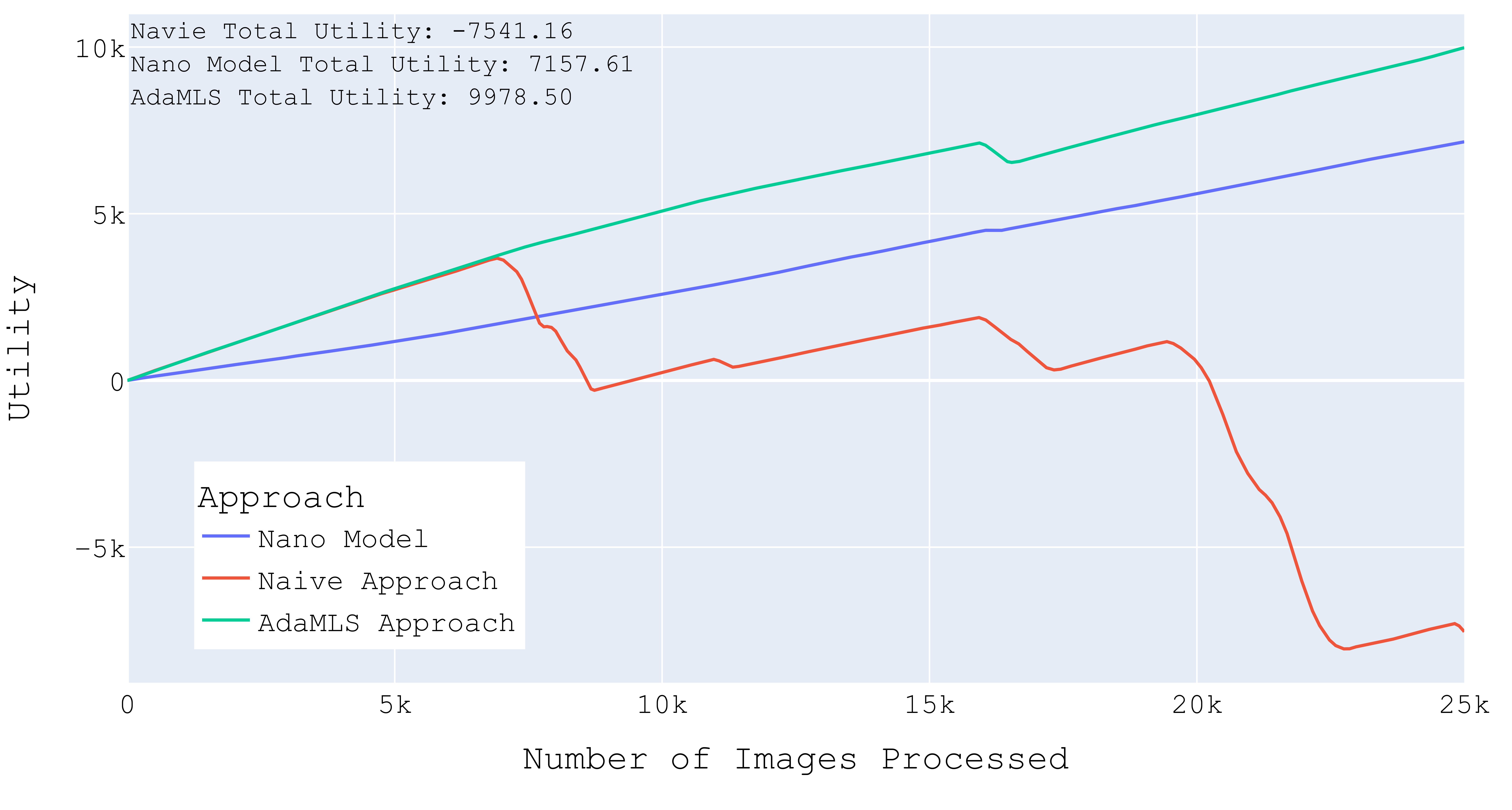}}
\caption{Utility Function Over Requests processed}
\label{fig3}
\end{figure}

\section{Related Work}
The inception of self-adaptive systems dates back to IBM's autonomic computing vision~\cite{sas_review}. The principle's application to ML Systems (MLS) is commenced by seminal work~\cite{garlan}, which doesn't consider model switching as an adaptive tactic despite studies on retraining ML components' cost-benefit trade-offs~\cite{maria2}. In addition, the concept of self-adaptation of ML was introduced as a primer in~\cite{henrykarthikml}. However, it wasn't realized or elaborated in detail. While Convolutional Neural Networks (CNNs) have enhanced object detection~\cite{cvpr2017}, the optimal architectural selection remains a real-world challenge. Recent advancements have primarily concentrated on enhancing individual models~\cite{cvpr2022,objanchor,objpro,objtext,objdrone}, neglecting system-wide adaptability. A recent survey summarizes the use of ML for engineering self-adaptive systems and it also highlights the underutilization of unsupervised learning~\cite{danny}.  AdaMLS, our solution, fills this gap by combining unsupervised learning and model switching to boost MLS adaptability.  and echoes the need for robust architectural practices for ML systems~\cite{deciframe}.

\section {Conclusion}
To conclude, we introduced AdaMLS, an innovative solution that engineers self-adaptive Machine Learning Systems (MLS) by employing unsupervised learning for dynamic model switching. Preliminary evaluations, based on an object detection system example indicate that AdaMLS can effectively mitigate run-time uncertainties and outperforms both traditional and standalone models. Thereby, offering significant QoS improvements. AdaMLS stands as a significant advancement, showcasing the potential of engineering MLS with self-adaptation capabilities. Importantly, it paves the way for MLS to execute seamless model switching to maintain optimal QoS under varying run-time uncertainties. Looking forward, we intend to explore a diverse range of learning techniques and model-switching strategies to further enhance the adaptability of AdaMLS. Emphasis will also be laid on broadening its applications to different domains, and on improving the environmental and economic sustainability of MLS, thereby revolutionizing the future of MLS implementation and design.

\bibliographystyle{IEEEtran}
\bibliography{references}

\end{document}